\documentclass[11pt]{article}
\usepackage{amsmath,graphicx}
\usepackage{amssymb, amsmath} 
\usepackage{blkarray}
\usepackage{mathtools}

\begin{document}
\title{Deriving diffeomorphism symmetry}
\author{
H.~B.~Nielsen$^a$, Astri Kleppe$^b$ \\[5mm]
\itshape{$^a$ The Niels Bohr Institute, Copenhagen, Denmark, hbech@nbi.dk}\\
\itshape{$^b$}SACT, Oslo, Norway, astri.snofrix@org}
\date{}
\maketitle                                       
\begin{abstract}
In an earlier article, we have "derived" space, as a part of the Random Dynamics project. In order to get locality we need to obtain reparametrization symmetry, or equivalently, diffeomorphism symmetry. 

There we sketched a procedure for how to get locality by first obtaining reparametrization symmetry, or equivalently, diffeomorphism symmetry. This is the object of the present article.
\end{abstract}

\section{Introduction}
In an earlier article \cite{space}, we have "derived" space, as a part of the Random Dynamics project \cite{rd}. 
Since we want to have locality, we also need to derive reparametrization symmetry, or more generally, diffeomorphism symmetry \cite{hbben}, essentially ensuring that the choice of coordinates plays no role in the formulation of the physical laws. 

We propose that diffeomorphism symmetry comes about as a result of a selection principle, in reality a selection principle for how Nature "chooses" its symmetry groups, a scheme that has been developed by Holger Bech Nielsen and his collaborators \cite{smallreps}. The initial idea was that the small representations of the Standard Model gauge group 
\begin{equation}  
SMG =  S(U(2) \times   U(3))
\end{equation}  
is a signature of such a selection principle, singling out groups that have the ``smallest'' representations.

In the present article we use similar arguments, but instead of taking the Standard Model group $SMG = S(U(2)  \times   U(3))$ as the selected group, we consider the combined diffemorphism-and-gauge group 
\begin{equation}
{\mathcal{B}}=\lbrace(\lambda,\varphi)\hspace{1mm}|\lambda \in {\mathcal{G}}, \varphi \in {\mathcal{D}}\rbrace
\end{equation}
where 
${\mathcal{G}}$ is the group of all gauge transformations that map the four-dimensional spacetime manifold ${\mathcal{M}}$ on the 12-dimensional manifold of $SMG$: the Lie group is a manifold, 
\[  
\lambda: {\mathcal{M}}  \rightarrow {SMG}
\]
and 
${\mathcal{D}}$ is the group of diffeomorphisms, a diffeomorphism $\varphi$ given by a bijective differentiable map
\[  
\varphi: {\mathcal{M}}  \rightarrow {\mathcal{M}}
\]

\subsection{An alternative to grand unified models}

A major part of the success of the GUT $SU(5)$ model is that the representations of the $SU(5)$ gauge group automatically represent the $SU(5)$ subgroup $S(U(2) \otimes U(3))$ with the Standard Model Lie algebra. The GUT $SU(5)$ group thus presents the needed restrictions on the allowed represenstations of the Standard Model algebra. 
Any successful GUT group, like for example $SO(10)$, reproduces the same restrictions as $SU(5)$ on the representations of the Standard Model Lie algebra, restrictions corresponding
to $S(U(2) \otimes U(3))$. 

Any viable alternative to the GUT scheme must thus supply a prediction not only of the Standard Model Lie algebra, but also of the group structure. 
There are however many possible scenarios, so unless one has some guiding principle for selection the unificaton group, there isn't much predictive power. 

One way to get the Standard Model without a GUT scheme, is by using some selection principle for how Nature selects the Standard Model group. The underlying philosophy is that of Random Dynamics, namely that the fundamental physics is random, and that the observed symmetries are emergent. If only some symmetries emerge, supposedly by accident, but maybe even by some more precise mechanism, then the initially random action could be considered as taking random values for some small region of the value space of the representation of the group, with the transformation properties of the fields or degrees of freedom under the group. The elements of a representation of the group in question then move quite slowly as the group elements themselves vary. (One can vary the group elements much before one varies the fields or matrices of the representation). The slower the representation moves as a function of the variation of the group elements the more likely it is that a symmetry emerges, since displacements inside the group itself corresponding to a small region (over which we assume essentially constancy of the action) become bigger with a slower representation motion rate. A symmetry of the random action is thus more likely to occur when the symmetry is represented by "slowly moving" representation elements (e.g. matrices). 

By means of some "goal quantities" we single out the groups that have the largest chance to emerge from a random action model, favouring the experimental gauge group and dimension of spacetime.

\subsection{Groups and algebras}

In Yang Mills theories, only the Lie algebra is important, since two groups ${\mathcal{G}}_1$ and ${\mathcal{G}}_2$ that have the same Lie algebra also have the same Yang Mills system.
There are however many Lie groups with the same algebra. These groups are locally similar, but globally they can be very different, with different representations. For example, $SU(2) \neq SO(3)$, as for $SU(2)$ we have $j = 0,1/2,1,3/2,\ldots$, while for $SO(3)$ $j = 0,1,2,\ldots$, and it is only by studying the representations of Nature like $q_L$, $q_R$, the Higgs and so on, that we can establish which groups are at stake. 
To a group corresponds
\begin{itemize}

\item The Lie algebra and thus the structure constants $f^l_{km}$ and the Yang-Mills Lagrangian ${\mathcal{L}}_{YM}$.

\item The system of allowed representations, a given set of representations only being allowed by some Lie groups.
\end{itemize}

The covering group (for a given Lie algebra) can manage all the representations, so the goal is to find the most choosy group, i.e. the one that allows the fewest representations - which also corresponds to experimental data.
Our point of deparure is the Standard Model Lie algebra
\begin{equation}
S(U(2) \times  U(3)) \sim R \times  SU(2) \times  SU(3)\sim U(1) \times  SU(2) \times  SU(3)
\end{equation}
and since we don't find all its possible representations in Naure, we will concentrate on the Lie group rather than on the Lie algebra. It is so to speak stronger to "predict" the group $S(U(2) \times  U(3))$, such that
\begin{equation}
             det   \left (\begin{array}{rcl}
                  .   &  .{\hspace{4mm}}0{\hspace{4mm}}0{\hspace{4mm}}0\nonumber\\
                  .   &  .{\hspace{4mm}}0{\hspace{4mm}}0{\hspace{4mm}}0\nonumber\\
                  0   &  0{\hspace{4mm}}.{\hspace{4mm}}.{\hspace{4mm}}.\nonumber\\
                  0   &  0{\hspace{4mm}}.{\hspace{4mm}}.{\hspace{4mm}}.\nonumber\\
                  0   &  0{\hspace{4mm}}.{\hspace{4mm}}.{\hspace{4mm}}.\nonumber\\
                      \end{array}
                \right) = 1,
\end{equation}
a group which admits all phenomenological representations, which all obey the rule  
\begin{equation}
\frac{y}{2}+j_3 + \frac{1}{3}"triality" = 0(mod1)
\end{equation}

\section{Skewness}
Small representations as one possible selection principle, 
but another way of singling out Nature's chosen group, is by studying group \textit{skewness} \cite{skew}, defined as a lack of symmetry.

Nature seems to select the Lorentz group with the smallest representations; perhaps space moreover prefers those dimensions that give the skewest Lorentz group. The Standard Model group $SMG$ = $S(U(2) \times  U(3))$ is very skew, and most probably very "complicated".   

There is always the worry that the choice of ``goal property'' is such that it gets dramatically bigger or smaller with the dimension or some other size parameter of the group. In the case of choosing a skewness measure, this can be dealt with by defining it as
\[
\frac{ln({\rm{ number\hspace{2mm} of\hspace{2mm} outer\hspace{2mm} automorphisms}})}{{\rm{rank\hspace{2mm} of\hspace{2mm} the\hspace{2mm} group}}}
\]

\subsection{Inner and outer automorphisms}
The degree of skewness is thus a function of the number of outer automorphisms of the group ${\mathcal{G}}$.
An automorphism is an isomorphism of the group onto itself, 
\begin{equation}
\beta: {\mathcal{G}} \rightarrow {\mathcal{G}}
\end{equation}
i.e. a correspondance $\phi$ of ${\mathcal {G }}$ with itself respecting the
group multiplication, and such that
$\phi$ is bijective and $\phi(gh)=\phi(g)\phi(h)$, $g, h \in {\mathcal{G}}$.

The map $\beta(g)$ is an inner automorphism if there is an element $h \in {\mathcal{G}}$, such that 
for all $g \in {\mathcal{G}}$, 
\begin{equation}
\beta(g)=\beta_h(g) = hgh^{-1}
\end{equation}
The group of outer automorphisms ${\mathcal{O}}$ is then defined modulo the inner automorphisms in the sense that in the group of all automorphisms ${\mathcal{A}}$, we discern the subgroup of inner automorphisms,
\begin{equation}
{\mathcal{A}}_{inn} = \lbrace\beta_h | h \in {\mathcal{G}}\rbrace,
\end{equation}
and then define the group of outer automorphisms as
\begin{equation}
{\mathcal{A}}_{out} = \lbrace{\mathcal{O}}/ \lbrace\beta_h | h \in {\mathcal{G}}\rbrace
\end{equation}

For the Standard Model group, we have that
\begin{itemize}
\item The automorphisms of ${\bf R}$ ($\sim$ the $U(1)$ factor) are scalings with a factor $k \neq 0$.

\item The $SU(2)$ factor has complex conjugation (in the defining representation) as an automorphism, it is however an inner automorphism.

\item For the $SU(3)$, as for all $SU(N)$ algebras with $N \geq 1$, complex conjugation is an outer automorphism.
\end{itemize}
All outer automorphisms of the Standard Model algebra are combinations of these, since an automorphism maps the three invariant subalgebras into three isomorphic invariant subalgebras. There are infinitely many such automorphisms, but the Standard Model algebra together with the set of Standard Model properties (the rule system) is invariant under only one outer automorphism, namely complex conjugation of the $SU(3)$ combined with the $U(1)$ scaling factor $k = -1$. 

Among all algebras of dimensionality up to 12 dimensions, taking quantization rule systems into account, there are four combinations of algebras and rule systems that have no generalized outer automorphisms, namely those with semisimple algebras $\mathfrak{su}(3)$ and $\mathfrak{so}(3)$.

\section{The size of a representation}
The other suggested selection principle, namely the size of a representation, was inspired by the fact that after the trivial representation, the lowest-dimensional non-abelian representations in the Standard Model are the remarkably small representations of $SU(2)$ and $SU(3)$.

A probablility argument for the presence of a selection principle can be formulated as follows:  
look at $S(U(2) \times  U(3))$ and count the Lie groups of similarly low rank \cite{RughSuhr}. It turns out that there are about $2^8$ = 256 groups with low dimension (up to 12, i.e. the dimension of the $SMG$). Among these about 256 groups, $S(U(2) \times  U(3))$ is singled out - most probably by means of some selection principle like the size of the representations. 

In order to obtain a more precise formulation of the selection principle, we need to establish what we mean by the ``size'' of a representation. For this purpose we define a measure for this size in terms of the quadratic Casimir operators, which 'tag' the representations in the sense that they are not defined for the algebra itself, but only for the representations. 

A general Casimir invariant is a function $f(F)$ of the Lie group generators $F_j$ which is invariant under the group and commutes with all the generators,
\[
\left [f(F),F_j \right ]=0.
\]
The generators $F_j$ of the group constitute a basis for the corresponding Lie algebra and satisfy the commutation relation $[F_i,F_k] = f^j_{ik}F_j,\hspace{3mm} i,k,j = 1,2,\ldots.,d_G$,
where $d_G$ is the dimension of the group and $f^j_{ik}$ are the structure constants by means of which we can construct a Killing metric tensor $g_{kl} = f^j_{ki}f^i_{jl}$.
The quadratic Casimir operator 
\begin{equation}\label{ett}
{\bf{C_2}} = g^{kl}F_kF_l
\end{equation}
is used for measuring the ``size'' of a representation $r$. This is done by normalizing the quadratic Casimir of the representation by dividing it with the quadratic Casimir for the adjoint representation, which consists of $d_G  \times  d_G$ matrices ${\bf{A}}_j$, such that $({\bf{A}}_j)^k_l = -f^k_{jl}$. The metric can thus be written $g_{kl} = Tr({\bf{A}}_k{\bf{A}}_l)$, and 
in the first approximation, the ``size'' of the representation $r$ is taken to be 
\begin{equation}
{\mathcal{S}}=\left (\frac{C_r}{C_A} \right),
\end{equation}
where $C_r$ and $C_A$ are the Casimirs for the representation $r$ and the adjoint representation, respectively. 

In the search for the groups with the smallest representations, we thus examine the quadratic Casimir operators, bearing in mind that the quadratic Casimir is well defined only for irreducible representations. 
Our goal is to show that the combined group ${\mathcal{B}}=\lbrace(\lambda,\varphi)|\lambda \in {\mathcal{G}}, \varphi \in {\mathcal{D}}\rbrace$ has a measure which is smaller than the Standard Model group measure,
\begin{equation}\label{casi}
{\left (\frac{C_r}{C_A}\right)}_{\mathcal{B}} < {\left (\frac{C_r}{C_A}\right)}_{SMG},
\end{equation}
since this is a way of necessitating the existence of the group of diffeomorphisms. 
When we talk about $SMG$, it should be noted that we actually have a $SMG$ in each point of spacetime, corresponding to a product of $SMG$'s: $SMG \times  SMG\ldots \times  SMG$, this product however has the same size measure as $SMG$ itself, i.e. $(C_r/C_A)_{{SMG \times  SMG\ldots}}$ = $(C_r/C_A)_{SMG}$. 

According to Schur's lemma \cite{schur}, in an irreducible representation, any operator that commutes with all the generators of the Lie algebra must be a multiple of the identity operator. Therefore $C_r = g^{kl}F_kF_l = c_r {\bf{1}}$,where ${\bf{1}}$ is the $d_r  \times  d_r$ identity matrix, and $c_r$ is a coefficient which only depends on the representation $r$, so we have
\begin{equation}
{\mathcal{S}}=\frac{c_r{\bf{1}}}{c_A{\bf{1}}}=\frac{c_r}{c_A}
\end{equation}
The point is to minimize the relation $c_{\rho}/c_A$, with some normalization of $c_A$ (the normalization in reality being arbitrary).

We are interested in the SU(N) group, which has the defining representation
\begin{equation}
              \left (\begin{array}{rcl}
                  b_1 \nonumber\\
                  
                 \vdots\\
                  b_N \nonumber\\
                      \end{array}
                \right) 
\end{equation}
and the group elements are ${\bf{U}}$ = NxN complex unitary matrices with determinant 1.
The matrices ${\bf{U}}$ are $\approx {\bf{1}}$, and can be written as ${\bf{U}} = e^{i{F}}$, with infinitesimal generators $F$. 
These $F's$ constitute a real vector space with the dimension $N^2 - 1$, i.e. the dimension of SU(N), and
we can choose a basis in the $F$-space, $F_1,F_2,\ldots,F_{N^2-1}$, which can be normalized. 

In an irreducible representation 
\[
\rho: {\mathcal{G}} \rightarrow (Matrices)
\]
\[
\rho(g) = \rho(1)+i\rho(F_j)g^j
\]
the quadratic Casimir $g_{kl}\rho(F_k)\rho(F_l)$ is only an eigenvalue, but it represents how intensively $\rho(g)$ varies, in the sense that a small $c_{\rho}$ corresponds to a "lazy" $\rho$.

The Casimirs thus function as a crude measure for how much the representation matrix varies as a function of the group element it represents. In a lattice context we take the contribution from one plaquette to be the trace of some representation of the group, the most general action $S_{\Box}$ is then a linear expansion on traces of all the possible representations of the gauge group, and the traces of the smallest representations supposedly dominate.   
This domination corresponds to the variation of the action as a function of how the combination of the link variables varies over the gauge group, and if the action varies relatively slowly over the group, it's taken as an indication that it also varies relatively slowly when we vary the gauge group.

So with an action which is dominated by the contributions from small representations, the variation along the gauge variation is presumeably quite small, a situation corresponding to small quadratic Casimir values. This increases the chance that an action which was not perceived as invariant under a gauge transformation, would nevertheless appear as gauge invariant.

To get an intuition of this "smallness" of a representation, consider SU(2) with its quadratic Casimir 
$\vec{J}^2$. On an irreducible representation, $\vec{J}^2$ effectively only takes one value, i.e. it has the same eigenvalue on the whole representation.  
With $\vec{J}^2 = g_{ij}J^iJ^j$ we have a notion of distance, and we can
visualise $\vec{J}^2$ as performing a rotation,
\[
\vec{J}^2|a> = j(j+1)|a>
\]
where the "smallness" of the representation means that $|a>$ is just slightly rotated,
 \begin{figure}[htb]
    \begin{center}
    \includegraphics{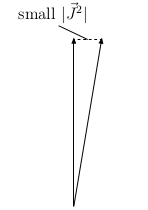}
\end{center}
    \end{figure}

By means of the Casimir measure we can thus define a size measure, making it meaningful to say that the representations of the non-abelian parts of the Standard Model are "small". In the abelian case, it is however problematic to establish what we mean by the "size" of a representation. We cannot apply a similar reasoning for the U(1) groups as for non-abelian groups, because in the abelian case we cannot use the dimension of the representation as a measure, since abelian groups always have 1-dimensional representations, so dimension doesn't tell anything. There simply is no Casimir element defined, since for an abelian Lie algebra the Killing form is zero.

What we can do is to consider the ratio of the charges of the representation and refer to a ``Quantum of charge'', for example the Millikan unit quantum. 
The unique abelian invariant subgroup in the Standard Model gauge group corresponds to the weak hypercharge. That can however not be used as the Quantum, since the quantum for $y/2$ is 1/6, while right-handed charged leptons have $y/2=-1$, which is 6 times larger than 1/6, and 6 is obviously not the smallest integer after zero.

So instead we consider non-invariant abelian subgroups, and define an abelian representation as small if it has relatively many charges (generators of the Lie algebra) with only the values 0, 1 or -1 measured in the Quantum.

It turns out that the Standard Model has a relatively large set of such charges, so in this perspective, even from the abelian point of view the Standard Model is a model with ``small'' representations.

\subsection{The size of a composite group}

In order to normalize the measure by means of the Casimir of the adjoint representation, we clearly need that there is a well-defined adjoint representation. In the case of a composed, non-simple group like ${\mathcal{B}}=\lbrace(\lambda,\varphi)|\lambda \in {\mathcal{G}},\varphi \in {\mathcal{D}}, \rbrace$, there is however no straightforward definition of the adjoint representation.
For ${\mathcal{B}}$, we therefore must find a way of varying the two adjoint normalizations relative to each other.    

In order to achieve this, we seek to establish a (faithful, 1-1) representation $r$ of ${\mathcal{B}}$ on which we define a metric, whereby the image of ${\mathcal{B}}$ becomes a manifold with a metric, allowing us to define a volume.

One way to do this is to establish a (faithful) representation $r$ of ${\mathcal{B}}$ on which we define a metric. Thus the image of ${\mathcal{B}}$ becomes a manifold with a metric, which makes it possible to define a volume. 
The measure $(c_F/c_A)_{\mathcal{B}}$ is then given as the volume ratio of the two representations, taken to the power $2/d_{{\mathcal{B}}}$,  
\[
\left(\frac{c_F}{c_A}\right)_{\mathcal{B}}=\left ( \frac{V_F}{V_{adj}}\right) ^{2/d_{{\mathcal{B}}}}
\] 
In the representation picture $c_r \sim g_{ik}$, i.e.
\begin{equation}
g^{({\mathcal{G}})}_{ik} = \frac{c_A}{c_r}g^{(r)}_{ik}
\end{equation}
thus
\begin{equation}
\frac{Vol({\mathcal{G}})}{Vol({\rm{representation \hspace{1mm}r}})} = \left (\frac{c_A}{c_r}\right )^{d_{G/2}}
\end{equation}
and in the case of
\begin{equation}
{\mathcal{G}} = {\mathcal{G}}_1  \times   {\mathcal{G}}_2  \times  \ldots \times   {\mathcal{G}}_k
\end{equation}
where the ${\mathcal{G}}_j$ are simple, the quantity
\begin{equation}\label{comp1}
\frac{c_r}{c_A} = \left[ \left(\frac{c_{r_1}}{c_{A_1}}\right )^{d_1}\left (\frac{c_{r_2}}{c_{A_2}}\right )^{d_2}\ldots\left (\frac{c_{r_k}}{c_{A_k}}\right )^{d_k}\right ]^{\frac{1}{d_1+d_2+\ldots+d_k}}
\end{equation}
is a "good quantity".

\subsection{Competing groups}
For an irreducible representation $r$, consisting of a set of $r  \times  r$ matrices ${\bf{M}}_j$, the second-order index $I_2(r)$ of the representation is defined by
\begin{equation}
Tr({\bf{M}}^i_r{\bf{M}}^j_r)=I_2(r)\delta^{ij}
\end{equation}
Taking the trace of (\ref{ett}), we get for the quadratic Casimir 
\begin{equation}
c_2(r)=I_2(r)d_G/d_r
\end{equation}
where $d_r$ is the dimension of the representation $r$, and $d_G$ is the group dimension.
For the defining, fundamental representation $N$ of $SU(N)$ (i.e., in reality the algebra ${\mathfrak{su}}(N)$) the second-order index is $I_2(N) = 1/2$, and for the adjoint representation $I_2(A) = N$ and $d_A = N^2-1$, which gives $\left (\frac{c_N}{c_A} \right)_{SU(N)} =\frac{N^2-1}{2N^2}$, thus for $SU(2)$ $\left (\frac{c_N}{c_A} \right)_{SU(2)}= 3/8$.

There are presumably other candidates, like $SO(N)$, with the fundamental representation consisting of $N  \times  N$ real matrices.
One can define higher tensor representations from the defining vector representation $N$, but there are also additional, double-valued spinor representations, similar to $SO(3) \sim SU(2)$, generated by direct products of the fundamental spinor. 

In the case of the $SU(N)$ group, the faithful representation with the smallest quadratic Casimir, is the fundamental representation $N$, while for the $SO(N)$ group the picture is much more complicated, as the faithful representation $F$ with the smallest quadratic Casimir might be either the vector representation, or the spinor representation, the spinor representation being the winner for $N < 8$.

For the vector representation, the second-order index for the $SO(N)$ fundamental and adjoint representations are $I_2(N) = 2$ and $I_2(A) = 2N-4$, respectively, and the dimension of the adjoint representation $d_A = N(N-1)/2$, thus
$\left (\frac{c_N}{c_A} \right)_{SO(N)}=\frac{N-1}{2(N-2)}$, and for the corresponding spinor representation we have $\left (\frac{c_N}{c_A} \right)_{SO(N)}^{spinor}=\frac{N-1}{2(N-2)}\frac{N}{8}$.
Another competitor is ${\mathfrak{sp}}(2N)$, with $\left (\frac{c_N}{c_A} \right)_{sp(2N)}=\frac{2N+1}{4(N+1)}$, thus
\begin{equation}
\begin{array}{rcl}
&\left (\frac{c_N}{c_A} \right)_{SO(N)}^{vector}  & =\frac{N-1}{2(N-2)},\hspace{8mm}
\left (\frac{c_N}{c_A} \right)_{SO(N)}^{spinor}    =\frac{N-1}{2(N-2)}\frac{N}{8}\\
&\left (\frac{c_N}{c_A}\right)_{SU(N)}          & =\frac{N^2-1}{2N^2},\hspace{10mm}\left (\frac{c_N}{c_A} \right)_{sp(2N)}   =\frac{2N+1}{4(N+1)}
\end{array}
\end{equation}
In the search for the groups chosen by Nature, we examine the $(c_N/c_A)$ for the different groups, but we also worry about possible differences between 3+1 and 4 spacetime dimensions.
For example, for dimension d = 3 +1, we have for the Lorentz group
$SO(3,1) \sim SL(2,C)$, while for d = 4, $SO(4) \sim SU(2)_R  \times   SU(2)_L$. For both d = 3 + 1 and d = 4, the Lorentz group however has the same small ${\mathcal{S}}$,
\begin{equation}
{\mathcal{S}}= \left (\frac{c_N}{c_A}\right) = \frac{\frac{1}{2}(1+\frac{1}{2})}{1(1+1)} = \frac{3}{8}
\end{equation}
while for d = 2 and d = 5, the value is bigger.

The Lorentz group $SO(d-p,p)$ in reality comes from a symmetric metric $g_{ik} = g_{ki}$. If $g_{ki}$ instead had been antisymmetric, we would have symplectic groups, which are not so competitive, as they have bigger $(c_N/c_A)$. 
We consider the Lorentz group as a function of the dimension $d$ and of the "geometry", in the sense of the dependence on whether $g_{ik}$ is symmetric, antisymmetric or nonexistent. But an antisymmetric $g_{ik}$ actually does very poorly, while for $d$ = 3,4 it looks good for symmetric $g_{ik}$; and for the case without metric for $d$ = 2 \cite{dim}.
 
Among simple groups, SU(2) has the smallest $(c_N/c_A)$, but in order to allow SU(3) be let in, some cooperation with SU(2) is necessary, since the Spin(5), which is the covering group of SO(5), in reality seems to outdo SU(3).

SU(3) however has a relatively big center $Z_3$, so if we divide by the group center SU(3) is in good shape, since the SO(5) covering group has a smaller center. For SU(3), the number of elements in the center is 3, while the center of SU(2) merely has 2 elements, and likewise for Spin(5). We thus redefine our measure of representations as
\begin{equation}\label{center}
{\mathcal{S}}=(\frac{c_r}{c_A})\frac{1}{\left [{\rm{(Number\hspace{1mm}of\hspace{1mm}elements\hspace{1mm}in\hspace{1mm}the\hspace{1mm}center)}}\right]^{2/d}}
\end{equation}
We in reality consider volumes:
\[
\frac{Vol(SU(3))}{Vol(SU(3)/Z_3)}=3
\]
And with $SU(2)/Z_2 = SO(3)$, 
\[\frac{Vol(SU(2))}{Vol(SO(3))} = 2.
\]
the quadratic Casimir being a sort of area in the group.

\section{The group of diffeomorphisms}
We define our group ${\mathcal{B}}$ as the combination of the gauge transformations of $SMG$ and the group of diffeomorphisms.  

A diffeomorphism so to say moves a function, by the operation
\[
x^{\mu}  \rightarrow  x^{\mu} + \eta^{\mu} 
\]
The displacement takes place in a given direction, and if we perceive the diffeomorphisms as vectors over a manifold, then for infinitesimal $\eta^{\mu}$ the set of displacements $\{\eta^{\mu}\}$ constitutes a tangent field. 
The group of diffeomorphisms does not have a (usual) Lie algebra, but we take as
the Lie algebra a set of fields $\lbrace\epsilon^{\mu} \rbrace$ corresponding to the tangents 
\begin{equation}
f(x) = \sum_{\mu}\epsilon^{\mu} \partial_{\mu},
\end{equation}
which amounts to substituting a manifold with a space of functions on the manifold,
\begin{equation}
[f_1(x),f_2(x)] = [\sum_{\mu}\epsilon^{\mu} \partial_{\mu},\sum_{\nu}\epsilon^{\nu} \partial_{\nu}],
\end{equation}
and then we could take
\begin{equation}
C = \int g_{\mu \nu}\epsilon^{\mu}(x)\epsilon^{\nu}(x)d\mathrm{x}
\end{equation}
as a kind of Casimir.
There are scarcely any outer automorphisms for the group of diffeomorphisms, and if all the automorphisms for the group of diffeomorphisms are inner, the group of diffeomorphisms is maximally skew.
It should however be noted that the group of diffeomorphisms depends on the topology of the space on which it is operating, for example for ${\bf{R}}^4$, the diffeomorphism group has a trivial center.

Even though the Lie algebra for the group of diffeomorphisms is not a usual Lie algebra, the group is still a Lie group. 
Consider the mappings of a manifold onto itself, ${\mathcal{M}}$, $\varphi: x \rightarrow x'$, i.e.
\[
\varphi: {\mathcal{M}}\rightarrow {\mathcal{M}} \hspace{4mm}   {\rm{where\hspace{2mm} \varphi \hspace{2mm} is}} 
\]
\noindent 
\begin{itemize}
\item bijective,

\item sufficiently many times continuously differentiable,

\item a group under the group of diffeomorphisms, 
\end{itemize}

then ${\varphi: {\mathcal{M}}\rightarrow {\mathcal{M}} }$ is really a "Lie group", which is clear by considering
$\varphi + \delta\varphi$ and take the commutators 
$[\delta\varphi_1,\delta\varphi_2] = [1+\delta\varphi_1,1+\delta\varphi_2] \neq 0$.

One difficulty we meet with respect to the combined group ${\mathcal{B}}$ is that the group ${\mathcal{D}}$ of diffeomorphisms is probably simple, while the group ${\mathcal{G}}$ of gauge transformations is not,
\[
 g \in {\mathcal{G}}| g: {\bf{R}}^4 \rightarrow SMG, 
\]
meaning
\[
g: {\bf{R}}^4 \rightarrow S(U(2) \times   U(3)); g(x) \in S(U(2) \times   U(3)), 
\]
and
\[
f,g: {\bf{R}}^4 \rightarrow S(U(2) \times   U(3)); (f g)(x) = f(x) \cdot g(x)
\]
and for a non-simple group we cannot define a straightforward measure like $(c_r/c_A)$ for the size of representation.
There is however one possibility to define a quadratic Casimir replacement, viz.
\begin{equation}
ln "c_N" = \int ln C \sqrt{g} \mathrm{d}^4x
\end{equation}
 where $g = det(g_{ik})$. 
The problem is that we cannot really have a metric, since a metric would not be diffeomorphism-symmetric. On the other hand, we don't quite need the metric $g_{ik}$, but only $\sqrt{g}$.

\section{The composite group}
Our selected group is ${\mathcal{B}}| \lbrace(\lambda,\varphi), \lambda \in {\mathcal{G}},\varphi \in {\mathcal{D}}\rbrace$, composed by the group of gauge transformations
\[
{\mathcal{G}} = \lbrace \lambda: {\mathcal{M}} \rightarrow SMG\rbrace
\]
and the group of diffeomorphisms 
\[
{\mathcal{D}} = \lbrace\varphi: {\mathcal{M}} \rightarrow {\mathcal{M}}\rbrace
\]

In order to investigate the group structure, we determine the action of the group elements.

Let $\Psi_l$ be a fermion state, and let $(\lambda,\varphi)$ operate on $\Psi_l$.
With the dfinition
\begin{equation}
(\lambda,\varphi) [\Psi_l](x) = \rho^k_l (\lambda(\varphi(x)))[\Psi_k](\varphi(x))
\end{equation}
where $\rho^k_l$ is the representation matrix, we want to determine the properties of the group operation $\circ$ of ${\mathcal{B}}$, i.e. of
$(\lambda_1,\varphi_1)\circ(\lambda_2,\varphi_2)$. 
First consider
\[
  \def\arraystretch{1.1}
  \begin{array}{r@{\;}l}
               (\lambda,1) [\Psi_k](x)& = \rho^l_k (\lambda(x))[\Psi_l](x)\\
     (1,\varphi) [\Psi_k](x) & = [\Psi_k](\varphi(x)) = [\Psi_k\circ \varphi ](x),\\
  \end{array}
\]
thus
\begin{equation}
  \def\arraystretch{1.1}
  \begin{array}{r@{\;}l}
(\lambda,1)\circ(1,\varphi)[\Psi_k](x) & =  (\lambda,1)[\Psi_k \circ \varphi](x)=\\  
                                       & =  \rho^l_k (\lambda(x))[\Psi_l(\varphi(x))]
  \end{array}
\end{equation}
Then consider
\[
  \def\arraystretch{1.1}
  \begin{array}{r@{\;}l}
(1,\varphi)\circ(\lambda,1)[\Psi_k](x) & =  (1,\varphi)\rho^l_k (\lambda(x))[\Psi_l(x)]=\\  
                                       & =  \rho^l_k (\lambda(\varphi(x)))[\Psi_l(\varphi(x))]
  \end{array}
\]
which leads to the conclusion that
 \begin{equation}
  \def\arraystretch{1.1}
  \begin{array}{r@{\;}l}
(\lambda,\varphi) & \neq (\lambda,1)\circ(1,\varphi)\\
(\lambda,\varphi) & = (1,\varphi)\circ(\lambda,1)
  \end{array}
\end{equation}
When investigating $(\lambda_1,\varphi_1)\circ(\lambda_2,\varphi_2)$ we use that
\[
(1,\varphi)[\chi](x) = (\chi \circ \varphi)(x)
\]
thus
\begin{equation}
  \def\arraystretch{1.1}
  \begin{array}{r@{\;}l}
(\lambda_1,\varphi_1)\circ(\lambda_2,\varphi_2)[\Psi_k](x) 
& = (\lambda_1,\varphi_1)\rho^l_k(\lambda_2(\varphi_2(x)))[\Psi_l(\varphi_2(x))]\\   
& = (\lambda_1,\varphi_1) [\rho^l_k(\lambda_2(\varphi_2(x)))(\Psi_l\circ\varphi_2)(x)]\\
& = \rho^m_l(\lambda_1(\varphi_1(x)))\rho^l_k(\lambda_2(\varphi_2\circ\varphi_1(x)))(\Psi_m\circ\varphi_2\circ\varphi_1)(x)\\
  \end{array}
\end{equation}
which we identify with
\begin{equation}
(\lambda_1,\varphi_1)\circ(\lambda_2,\varphi_2)[\Psi_k](x) = (\lambda_3,\varphi_3)[\Psi_k](x) =
\rho^m_k(\lambda_3(\varphi_3(x)))(\Psi_m\circ \varphi_3(x))\\   
\end{equation}
thus 
\begin{equation}
\varphi_3 = \varphi_2\circ\varphi_1,
\end{equation}
We demand that for all $[\Psi_k](x)$
\[
\rho^m_k(\lambda_3(\varphi_3(x)))= \rho^m_l(\lambda_1(\varphi_1(x)))\rho^l_k(\lambda_2(\varphi_2\circ\varphi_1(x)))
\]
which can only be achieved for a faithful representation, and 
\[
\lambda_1(\varphi_1) \cdot \lambda_2(\varphi_3)= \lambda_3(\varphi_3)
\]
(where $\cdot$ is the group operation for ${\mathcal{G}}$) which applies to all $x$, a special case being $\varphi_3^{-1}(x)$.
We perform the substitution $x \rightarrow \varphi_3^{-1}(x)$, thus obtaining
\[
\lambda_1(\varphi_1\circ \varphi_3^{-1}(x)) \cdot \lambda_2(x)= \lambda_3(x)
\]
and with $\varphi_3=\varphi_2 \circ \varphi_1$, we get $\varphi_1\circ \varphi_3^{-1}= \varphi_2^{-1}$ and
 \begin{equation}\label{lambda3}
\lambda_1(\varphi_2^{-1}(x)) \cdot \lambda_2(x)= \lambda_3(x),
\end{equation}
thus
\begin{equation}\label{comb}
(\lambda_1,\varphi_1)\circ(\lambda_2,\varphi_2)=(\lambda_3,\varphi_3)=(\lambda_1(\varphi_2^{-1}(\cdot)) \cdot \lambda_2,\varphi_2 \circ \varphi_1)
\end{equation}
Now $\lambda_1(\varphi_2^{-1}(\cdot)) \in {\mathcal{G}}$, and 
\[
(\lambda_1(\varphi_2^{-1}(\cdot),1))\Psi_k(x) 
=\rho^l_k(\lambda_1(\varphi_2^{-1}(x)))\Psi_l(x), 
\]
but exchange of argument in $\lambda$, i.e. 
$\lambda(x) \rightarrow \lambda(\varphi(x))$, is an automorphism in ${\mathcal{G}}$, and in this sense,
\[
\lambda_1(\varphi_2^{-1}(x))=[\Phi_{\varphi_2^{-1}}(\lambda_1)](x),
\]
is an automorphism in ${\mathcal{G}}$.
The product of two elements of ${\mathcal{B}}$ finally reads
\begin{equation}\label{comb}
(\lambda_1,\varphi_1)\circ(\lambda_2,\varphi_2)=(\lambda_3,\varphi_3)=(\Phi_{\varphi_2^{-1}}(\lambda_1)\cdot\lambda_2,\varphi_2 \circ \varphi_1)
\end{equation}
With the alternative definition $(\lambda,\varphi) [\Psi_l](x) = \rho^k_l (\lambda(\varphi^{-1}(x)))[\Psi_k](\varphi^{-1}(x))$, we moreover get that 
 \begin{equation}\label{coma}
(\lambda_1,\varphi_1)\circ(\lambda_2,\varphi_2)=(\lambda_3,\varphi_3)=(\Phi_{\varphi_2}(\lambda_1)\cdot \lambda_2,\varphi_1 \circ \varphi_2)
\end{equation}

\subsection{Subgroups of ${\mathcal{B}}$}

Does ${\mathcal{B}}$ = $\lbrace(\lambda,\varphi)\rbrace$ have any subgroups?
The relation (\ref{lambda3}) seems to indicate that
the gauge group ${\mathcal{G}}$ is a normal (invariant) subgroup of ${\mathcal{B}}$, which means that for $ b \in{\mathcal{B}}$, $b{\mathcal{G}}b^{-1} \subseteq {\mathcal{G}}$.

With $(\lambda,\varphi) [\Psi_l](x) = \rho^k_l (\lambda(\varphi(x)))[\Psi_k](\varphi(x))$ and (\ref{comb}), i.e.
\[
\lambda_3=\Phi_{\varphi_2^{-1}}(\lambda_1)\cdot \lambda_2,\hspace{1cm}\varphi_3 = \varphi_2\circ\varphi_1
\]
we take
\begin{equation}
  \def\arraystretch{1.1}
  \begin{array}{r@{\;}l}
(\Lambda,\varphi)\circ(\lambda,1)\circ(\Lambda,\varphi)^{-1}&=(\Lambda,\varphi)\circ(\lambda,1) \circ (\Phi_{\varphi}^{-1}(\Lambda),\varphi^{-1})=\\
                                                           & = (\Phi_{\varphi}(\Lambda)\cdot\Phi_{\varphi}(\lambda)\cdot\Phi_{\varphi}^{-1}(\Lambda),1) \in {\mathcal{G}}\\
  \end{array}
\end{equation}
and specifically for $(1,\varphi)$, we get $(1,\varphi)\circ(\lambda,1)\circ(1,\varphi)^{-1}=(\lambda,1)$, so for these specific representatives $(1,\varphi)$ of the cosets of 
\[
\lbrace(\lambda,1)|\lambda: \mathcal{M} \rightarrow  {\mathcal{G}}\rbrace,
\]
$(\lambda,1)$ is similarity transformation invariant, and we conclude that ${\mathcal{G}}$ is a normal subgroup of ${\mathcal{B}}$. This implies that ${\mathcal{B}}$ is not simple, but a semi-direct product group,
unless the subgroup ${\mathcal{D}}$ of diffeomorphisms also is normal, i.e.
\[
(\lambda,1)(1,\varphi)(\lambda,1)^{-1} \in {\mathcal{D}}
\]
Again using (\ref{comb}), we get that 
\begin{equation}\label{sub}
(\lambda,\omega)\circ(1,\varphi)\circ(\lambda,\omega)^{-1} = (\lambda,\omega)\circ(1,\varphi)\circ (\Phi_{\omega}^{-1}(\lambda),\omega^{-1})
= (\Phi_{\varphi}^{-1}(\lambda)\cdot\lambda^{-1},\varphi) \notin {\mathcal{D}}
\end{equation}
thus ${\mathcal{D}}$ is not an invariant subgroup of ${\mathcal{B}}$, and ${\mathcal{B}}$ is a semidirect product of ${\mathcal{G}}$ and ${\mathcal{D}}$,
\begin{equation}
{\mathcal{B}} = {\mathcal{G}} \rtimes {\mathcal{D}}
\end{equation}
This means that $\phi_{\varphi^{-1}}$ in (\ref{comb}) is a group homomorphism $\phi_{\varphi^{-1}}$ : ${\mathcal{D}} \rightarrow Aut({\mathcal{G}})$, where $Aut({\mathcal{D}})$ denotes the group of automorphisms of ${\mathcal{D}}$.

\section{To evaluate the size of ${\mathcal{B}}$}
We started out from the Standard Model group $SMG$, which in itself is a compact, 12-dimensional manifold.
When we go to the bigger group ${\mathcal{B}}$ encompassing the group of gauge transformations extended with the group of diffeomorphims, we are dealing with an infinite dimensional Lie group.
But this does not necessarily have to be so devastating, keeping in mind that
the effect of ${\mathcal{D}}$ in ${\mathcal{B}}$ in reality is nothing more than to dislocate the different $SMG$ in the product $SMG \times   SMG\ldots \times   SMG$, $g^{ab}\rho(F_a)\rho(F_b)$.
Compared to $\prod SMG$, the group ${\mathcal{B}}$ (where also ${\mathcal{D}}$ is included) will still have the same representations.

In deciding on how to measure the size of a representation, we have encountered a set of problems,
\begin{itemize} 
\item How to establish a viable 'size' for the $U(1)$ group in $SMG$.
\item How do we handle the problem with the adjoint representation in the case of ${\mathcal{B}}$?
\item How do we define $c_F/c_A$ for a semidirect product?
\end{itemize} 
In spite of all these problems, let us make an attempt to evaluate the difference between the measures for $SMG$ and ${\mathcal{B}}$, respectively.
Represent ${\mathcal{D}}$ by the Lorentz group, taken as $SO(3,1)$ or $SO(4)$, supposing we are in 3+1 or 4 dimensions, and use $(c_F/c_A)=3/8$ (with the corresponding group dimension 6). In accordance with (\ref{comp1}) we then define a tentative measure for the composite group ${\mathcal{B}}={\mathcal{G}} \rtimes {\mathcal{D}}$, as 
\begin{equation}\label{casim}
\begin{array}{rcl}
&''{\mathcal{S}}_{{\mathcal{B}}}''=\left[ \left(\frac{c_F}{c_A}\right)_{SU(2)}^{d(SU(2))}\cdot\left(\frac{c_F}{c_A}\right)_{SU(3)}^{d(SU(3))}\cdot\left(\frac{c_F}{c_A}\right)_{SO(4)}^{d(SO(4))}\right]^{\frac{1}{\sum d_i}}=\\
&=\left[\left(\frac{3}{8}\right)^3\cdot\left(\frac{4}{9}\right)^8\cdot\left(\frac{3}{8}\right)^6 \right]^{\frac{1}{17}}= 0.406213...
\end{array}
\end{equation}
where we for ${\mathcal{S}}_{\mathcal{G}}$ have used ${\mathcal{S}}_{SU(2)\otimes SU(3)}$, keeping in mind that the quadratic Casimir for $\prod SMG$ is the same as for $SMG$ itself, and ignored $U(1)$.

For $SMG$ alone, we get
\begin{equation}
\begin{array}{rcl}
&''{\mathcal{S}}_{{\mathcal{G}}}''=\left[ \left(\frac{c_F}{c_A}\right)_{SU(2)}^{d(SU(2))}\cdot\left(\frac{c_F}{c_A}\right)_{SU(3)}^{d(SU(3))}\right]^{\frac{1}{\sum d_i}}=\\
&= \left[\left(\frac{3}{8}\right)^3\cdot\left(\frac{4}{9}\right)^8\right]^{\frac{1}{11}}= 0.424320...,
\end{array}
\end{equation}
so in this crude approach, $''{\mathcal{S}}_{{\mathcal{B}}}'' < {\mathcal{''S}}_{{\mathcal{G}}}''$.
\begin{equation}
\begin{array}{rcl}
&''{\mathcal{S}}_{{\mathcal{G}}}''=\left[ \left(\frac{c_F}{c_A}\right)_{SU(2)}^{d(SU(2))}\cdot\left(\frac{c_F}{c_A}\right)_{SU(3)}^{d(SU(3))}\right]^{\frac{1}{\sum d_i}}=\\
&= \left[\left(\frac{3}{8}\right)^3\cdot\left(\frac{4}{9}\right)^8\right]^{\frac{1}{11}}= 0.424320...,
\end{array}
\end{equation}
There is however another aspect to this. Let us make no assumption about the dimension $N$ in $SO(N)$, and simply plot the expression (\ref{casim}) for $''{\mathcal{S}}_{{\mathcal{B}}}''$ as a function of $N$ for $N \leq 8$,
\begin{equation}
\begin{array}{rcl}
&''{\mathcal{S}}_{{\mathcal{B}}}''=\left[ \left(\frac{c_F}{c_A}\right)_{SU(2)}^{d(SU(2))}\cdot\left(\frac{c_F}{c_A}\right)_{SU(3)}^{d(SU(3))}\cdot\left(\frac{c_F}{c_A}\right)_{SO(N)}^{d(SO(N))}\right]^{\frac{1}{\sum d_i}}=\\
&=\left [\left( \frac{3}{8}\right)^3\cdot\left(\frac{4}{9}\right)^8\cdot \left(\frac{N(N-1)}{16(N-2)} \right)^{\frac{N(N-1)}{2}}\right]^{\frac{1}{11+\frac{N(N-1)}{2}}},\\
\end{array}
\end{equation}

 \begin{figure}[htb]
    \begin{center}
    \includegraphics[scale=0.70]{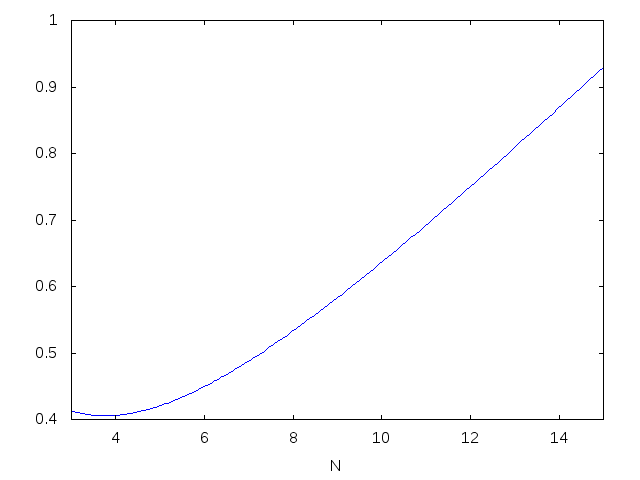}
\end{center}
    \end{figure}

corresponding to a minimal value for the size $''{\mathcal{S}}_{{\mathcal{B}}}''$ at $N=4$. 
These encouraging results, both for the relative smallness of $''{\mathcal{S}}_{{\mathcal{B}}}''$ compared to $''{\mathcal{S}}_{{\mathcal{G}}}''$, and well as the singling out of $N=4$, are of course based on a coarse evaluation, which is to be refined with a more precise formulation of the quadratic Casimir for the group ${{\mathcal{B}}}$, in order to accomplish a fair comparison between the sizes of the two groups.

\section{Conclusion}

In this article we have taken the first steps in "deriving" diffeomorphism symmetry, which is called for within the framework of the derivation of space. 
We have discussed different "goal quantities", especially the size of a representation of a group, identified as the size of the quadratic Casimir, which is connected with natural metric on the space of unitary matrices in the representations.

With this "goal quantity" in mind, we argue that diffeomorphism symmetry necessarily comes about, because the size of the bigger group, which is the semidirect product of the Standard Model group and the group of diffeomorphisms, is smaller than the size of th Standard Model group.

The next step will be to calculate the Casimirs for the entire group ${\mathcal{B}}$, and more precisely evaluate the size of ${\mathcal{B}}$ as compared with the size of the Standard Model group.

\end{document}